\def\apj{\rm ApJ}
\def\apjl{\rm ApJL}

\def\aj{\rm AJ}
\def\mnras{\rm MNRAS}

\documentclass{iaus}
\usepackage{graphicx}

\title[The Faint Structure of Galaxies] %% give here short title %%
{The Structure of Galaxies at Faint Light Levels: Probing Galaxy Assembly}

\author[Annette M. N. Ferguson]   %% give here short author list %%
{Annette M. N. Ferguson$^1$}

\affiliation{$^1$Institute for Astronomy, University of Edinburgh, Blackford Hill, Edinburgh UK EH9 3HJ \break email:ferguson@roe.ac.uk}

\pubyear{2006}
\volume{235}  %% insert here IAU Symposium No.
\pagerange{XXX}
\date{?? and in revised form ??}
\jname{Galaxy Evolution across the Hubble Time}
\editors{F.Combes \& J. Palous, eds.}
\begin{document}

\maketitle
\begin{abstract}
  Many clues about the galaxy assembly process lurk in the faint outer
  regions of galaxies.  Although quantitative study of these parts has
  been severely limited in the past, breakthroughs are now being made
  thanks to the combination of wide-area star counts, deep HST imagery
  and 8-m class spectroscopy. I highlight here some recent progress
  made on deciphering the fossil record encoded in the outskirts of
  our nearest large neighbours, M31 and M33.  
\end{abstract}

\firstsection % if your document starts with a section,

\section{The Faint Outer Regions of Galaxies: Motivation and History}

The study of the faint outskirts of galaxies has become increasingly
important in recent years. From a theoretical perspective, it has been
realised that many important clues about the galaxy assembly process
should lie buried in these parts. Cosmological simulations of disk
galaxy formation incorporating baryons have been carried out by
several groups and yield predictions for the large-scale structure and
stellar content at large radii -- for example, the abundance and
nature of stellar substructure and the ubiquity, structure and content
of stellar halos and thick disks. These models generally predict a
wealth of (sub)structure at levels of $\mu_V\sim 30$ mag/$\square''$
and fainter; their verification thus requires imagery and spectroscopy
of galaxies to ultra-faint surface brightness levels.

Since Malin first applied his photographic stacking and amplification
technique (e.g. Malin et al. 1983), it has been known that some
galaxies possess unusual low surface brightness (LSB) structures --
shells, loops, asymmetric envelopes -- in their outer regions.
Although limited to $\mu_B\lesssim 28$ mag/$\square''$, these images
were sufficient to demonstrate that even the most ``normal'' nearby
galaxies could become very abnormal when viewed at faint light 
levels (e.g. Weil et al 1997). Follow-up study of Malin's LSB
features, and several more recently-discovered examples (e.g. Shang et
al. 1999), has been severely limited due to the technical difficulties
associated with detecting and quantifying diffuse light at surface
brightnesses $\sim10$ magnitudes below sky.  The currently most
viable technique to probe the very low surface brightness regions of
galaxies is that of resolved star counts (e.g. Pritchet \& van den
Bergh 1994); I review here some recent results from studies of the
faint outer regions of M31 and M33.

\section{Substructure Around Galaxies}
\begin{figure}
 \includegraphics[height=8cm]{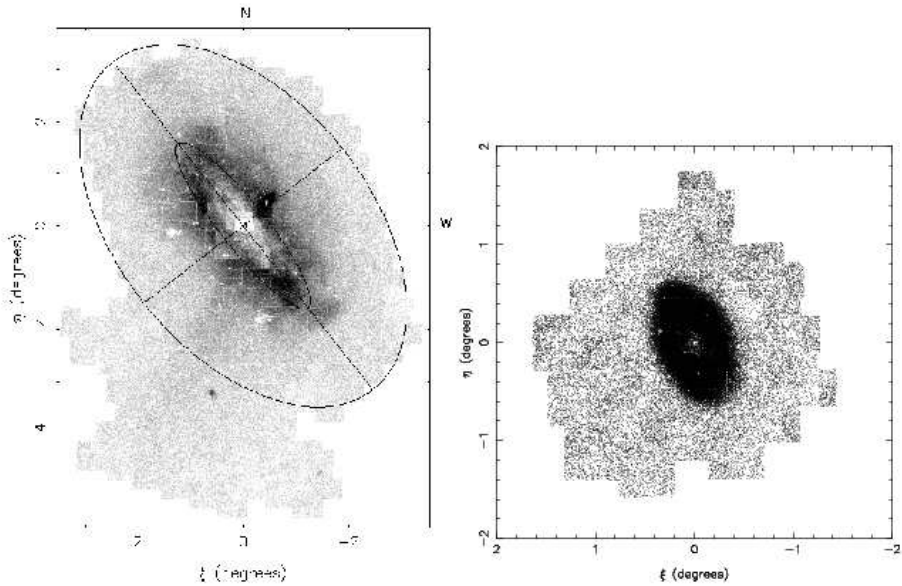}
  \caption{INT/WFC RGB star count maps of M31 (left) and M33 (right). These maps
span $125\times95$~kpc$^2$ and $56\times56$~kpc$^2$ on a side respectively. }
\end{figure}
Figure 1 shows maps of the red giant branch (RGB) population (age
$>1-2$~Gyr) in M31 and M33, covering 40 and 7 square degrees
respectively. Although the maps reach the same limiting depth
($\approx 3$ magnitudes below the tip of the RGB), it can be clearly
seen that outer regions of M31 contain a wealth of substructure that
is not present in M33 (Ferguson et al. 2002, Ferguson et al. 2007, in
prep). The faintest features visible by eye in the M31 map have
effective V-band surface brightnesses of $\mu_V\sim28-31$
mag/$\square''$; if substructure is present in M33, it must be of 
lower surface brightness than this.

Various features can be seen around M31, including a giant
stream in the south-east, stellar overdensities at large radii along
the major axis, a diffuse structure in the north-east and a loop of
stars projected near NGC~205.  It is of obvious importance to
establish the origin of this substructure.  Bullock \&
Johnston (2005) predict substructure in the outer regions of
galaxies due to tidal debris from the accretion and disruption of an
expected population of $\sim100-200$ luminous satellites.  Gauthier et al.
(2006) and Kazantzidis et al. (in prep) further show how the accretion
of such a population will heat and restructure the thin
disk, producing additional ``low latitude'' debris.  I
summarize here our current understanding of the substructure seen
around M31:
\begin{itemize}

\item{Deep HST/ACS CMDs reaching well below the horizontal branch have
    been obtained for 8 regions in the far outskirts of M31, including
    seven regions of visible substructure. Many of the CMDs exhibit
    different morphologies, suggesting variations in the mean age and
    metallicity of the constituent stars (Ferguson et al. 2005). 
    In all cases, the metallicity inferred from the RGB
    colour is significantly higher ([Fe/H]$\gtrsim-0.7$) than that of
    typical present-day low mass Local Group dwarf satellites
    ([Fe/H]$\lesssim-1.5$), implying that such objects are unlikely
    to be the progenitors of the observed substructure.  }
\item{All substructure fields lying near the major axis contain stars
    of age $\lesssim 1.5$~Gyr (Faria et al 2007; Richardson et al., in
    preparation); this includes the NE clump field, which lies at a
    projected radius of 40~kpc. Additionally, the major axis
    substructure fields are dominated by a strong rotational
    signature, similar to the HI disk, with a modest velocity
    dispersion (Ibata et al. 2005). Taken together, these results
    suggest that the ``low-latitude'' structure in M31 results
    primarily from perturbed outer thin disk.}
\item{The combination of line-of-sight distances and radial velocities
    for stars at various locations along the giant stream constrains
    the progenitor orbit (e.g.  Ibata et al. 2004; Fardal et
    al. 2006).  Currently-favoured orbits do not easily connect the
    more luminous inner satellites (e.g. M32, NGC~205) to the stream
    however this finding leaves some remarkable coincidences (e.g. the
    projected alignment on the sky, similar metallicities) yet
    unexplained.  The giant stream is linked to another overdensity,
    the diffuse feature lying north-east of M31's centre, on the basis
    of nearly identical CMD morphologies and RGB luminosity functions;
    orbit calculations suggest this connection is indeed likely. }
\end{itemize}  

\section{Diffuse Stellar Halos}

Tidal debris from accretion events will disperse over time and streams
from ancient events will have long merged to produce a diffuse stellar halo
(e.g. Abadi et al. 2005; Bullock \& Johnston et al 2005).  If
hierarchical growth is the main mode of mass assembly then such 
stellar halos should be a generic feature of galaxies. Unfortunately, 
there are very few observational constraints on the nature and ubiquity of 
stellar halos around galaxies (see however Zibetti et al. 2004). 
Even our understanding of the M31 and M33 stellar halos was
extremely poor until recently. I summarize some key results here:
\subsection{M31}
Using data from the INT/WFC survey, Irwin et al. (2005) combined
diffuse light surface photometry with resolved star counts to probe
the minor axis profile of M31 to a radius of $\sim55$kpc (effective
$\mu_V\sim32$~mag/$\square"$).  The profile shows an unexpected
flattening (relative to the inner R$^{1/4}$ decline) at a radius of
$\sim30$kpc, beyond which it can be described by shallow power-law
(index $\approx-2.3$), possibly extending out to 150~kpc (Kalirai et
al. 2006).  This compares favourably with the Milky Way halo, which
exhibits a power-law index of $-3.1$ in volume density (Vivas et
al. 2006).  The discovery of a power-law component which dominates the
light at large radii in M31 has profound implications for the
interpretation of all prior studies of the M31 ``stellar halo''. Since
these studies generally targetted regions lying within 30~kpc along
the minor axis, they most likely probed the extended disk/bulge region
of the galaxy and not the true stellar halo.

Keck/DEIMOS spectroscopy has been used to study the kinematics and
metallicities of stars in the far outer regions of M31. By windowing out
the stars which corotate with the HI disk, Chapman et
al. (2006) have detected an underlying metal-poor
([Fe/H]$\sim-1.4$), pressure-supported ($\sigma\sim100$~km/s)
component (see also Kalirai et al. 2006).  Although it has yet to be
proven that this component is the same one which dominates
the power-law profile at very large radius, the evidence is highly
suggestive.  Despite many years of thinking otherwise, it
thus appears that M31 does indeed have a stellar halo 
which resembles that of the Milky Way in terms of 
structure, metallicity and kinematics.

\subsection{M33}
Studies of the M33 stellar halo have also had a somewhat checkered history.
Mould \& Kristian (1986) measured [M/H]$\sim-2.2$ in a field located
at 7~kpc along the minor axis of M33 and, for the better part of two
decades, this metallicity was generally assumed to reflect that of the
M33 halo.  Recent work has found a significantly higher metallicity
for stars in this same field and, at the same time, suggested that the
field is actually dominated by the outer disk and not the stellar halo
(e.g. Tiede et al. 2004, Ferguson et al. 2007 in prep).  The detection
of a power-law structural component in the outskirts of M33 has so far
proved elusive, although the RGB clearly becomes narrower and more
metal-poor in these parts (Ferguson et al. 2007 in prep).  A recent
Keck/DEIMOS spectroscopic study has targetted two fields in the
outskirts of the galaxy, located at $\sim 9$~kpc along the major axis
(McConnachie et al. 2006).  Although the dominant kinematic component
in these fields exhibits strong rotation, there is tentative evidence
for an additional low-level metal-poor ([Fe/H]$\sim -1.5$) component
centered at the systemic velocity.  While further work is clearly
needed to elucidate the nature of this structure, the estimated
velocity dispersion ($\sigma \sim 50$~km/s) supports an association
with M33's true stellar halo.

\section{Future Outlook}
Quantitative study of the faint outskirts of galaxies provides
important insight into the galaxy assembly process. Much
recent work has focused on our nearest large neighbours, M31 and M33. While
both appear to show evidence for a metal-poor, pressure-supported
stellar halo (similar to that of the Milky Way), only M31 shows
evidence for recent accretion.  In order to put these results
in context, it is necessary to establish the properties of 
a larger sample of galaxies, spanning a
range in both host galaxy luminosity and Hubble type. With 8-m
class telescopes, it is possible to map the RGB populations in
the outskirts of galaxies to distances of $\lesssim 5$~Mpc; with
HST, this work can feasibly be extended to
$\gtrsim 10$~Mpc. Spectroscopic characterization of resolved
RGB populations beyond the Local Group is far harder and must
await the arrival of 30-m class telescopes.  

\begin{acknowledgments}
  I thank Scott Chapman, Daniel Faria, Rodrigo Ibata, Mike Irwin,
  Rachel Johnson, Kathryn Johnston, Geraint Lewis, Alan McConnachie
  and Jenny Richardson for their collaboration. Support from a Marie
  Curie Excellence Grant is acknowledged.
\end{acknowledgments}

%\begin{discussion}
%\end{discussion}


\begin{thebibliography}{}
\bibitem[Abadi et al.(2006)]{2006MNRAS.365..747A} Abadi, M.~G., Navarro, 
J.~F., \& Steinmetz, M.\ 2006, \mnras, 365, 747 
\bibitem[Bullock \& Johnston(2005)]{} Bullock, J.~S., \& 
Johnston, K.~V.\ 2005, \apj, 635, 931
\bibitem[Chapman et al.(2006)]{} Chapman, S.~C., Ibata, 
R., Lewis, G.~F., Ferguson, A.~M.~N., Irwin, M., McConnachie, A., \& 
Tanvir, N.\ 2006, astro-ph/0602604
\bibitem[Fardal et al.(2006)]{} Fardal, M.~A., Babul, 
A., Geehan, J.~J., \& Guhathakurta, P.\ 2006, \mnras, 366, 1012 
\bibitem[Faria et al.(2007)]{} Faria, D., Johnson, R., Ferguson, A.~M.~N., 
Irwin, M.~J., Ibata, R.~A., Johnston, K., Lewis, G.~F., \& Tanvir, N.~R. \ 2007, submitted
\bibitem[Ferguson et al.(2002)]{ferg02} Ferguson, A.~M.~N., 
Irwin, M.~J., Ibata, R.~A., Lewis, G.~F., \& Tanvir, N.~R.\ 2002, AJ, 124, 
1452  
\bibitem[Ferguson et al.(2005)]{ferg05} Ferguson, A.~M.~N., 
Johnson, R.~A., Faria, D.~C., Irwin, M.~J., Ibata, R.~A., Johnston, K.~V., 
Lewis, G.~F., \& Tanvir, N.~R.\ 2005, ApJL, 622, L109 
\bibitem[Gauthier et al.(2006)]{} Gauthier, J.-R., 
Dubinski, J., \& Widrow, L.~M.\ 2006, astro-ph/0606015 
\bibitem[Ibata et al.(2004)]{ibata04} Ibata, R., Chapman, S., 
Ferguson, A.~M.~N., Irwin, M., Lewis, G., \& McConnachie, A.\ 2004, MNRAS, 
351, 117 
 \bibitem[Ibata et al.(2005)]{ibata05} Ibata, R., Chapman, S., 
Ferguson, A.~M.~N., Lewis, G., Irwin, M., \& Tanvir, N.\ 2005, ApJ, 634, 
287 
\bibitem[Irwin et al.(2005)]{irwin05} Irwin, M.~J., Ferguson, 
A.~M.~N., Ibata, R.~A., Lewis, G.~F., \& Tanvir, N.~R.\ 2005, ApJL, 628, 
L105
\bibitem[Kalirai et al.(2006)]{} Kalirai, J.~S., et al.\ 
2006, \apj, 648, 389  
\bibitem[Malin et al.(1983)]{} Malin, D.~F., Quinn, 
P.~J., \& Graham, J.~A.\ 1983, \apjl, 272, L5 
\bibitem[McConnachie et al.(2006)]{} McConnachie, A.~W., 
Chapman, S.~C., Ibata, R.~A., Ferguson, A.~M.~N., Irwin, M.~J., Lewis, 
G.~F., Tanvir, N.~R., \& Martin, N.\ 2006, \apjl, 647, L25 
\bibitem[Mould \& Kristian(1986)]{} Mould, J., \& 
Kristian, J.\ 1986, \apj, 305, 591 
\bibitem[Pritchet \& van den Bergh(1994)]{pvdb94} Pritchet, 
C.~J., \& van den Bergh, S.\ 1994, \aj, 107, 1730
\bibitem[Shang et al.(1998)]{} Shang, Z., et al.\ 1998, 
\apjl, 504, L23 
\bibitem[Tiede et al.(2004)]{} Tiede, G.~P., Sarajedini, 
A., \& Barker, M.~K.\ 2004, \aj, 128, 224 
\bibitem[Vivas \& Zinn(2006)]{} Vivas, A.~K., \& Zinn, 
R.\ 2006, \aj, 132, 714 
\bibitem[Weil et al.(1997)]{} Weil, M.~L., 
Bland-Hawthorn, J., \& Malin, D.~F.\ 1997, \apj, 490, 664 
\bibitem[Zibetti et al.(2004)]{2004MNRAS.347..556Z} Zibetti, S., White, 
S.~D.~M., \& Brinkmann, J.\ 2004, \mnras, 347, 556 
\end{thebibliography}
\end{document}